\documentclass[reprint, amsmath,amssymb, showkeys, aps]{revtex4-1}

\usepackage{graphicx}
\usepackage{dcolumn}
\usepackage{bm}

\newcommand{\Comment}[1]{{}}
\usepackage{amsfonts,amsthm,amsmath,amssymb,slashed}
\usepackage[textwidth = 520 pt, textheight = 630 pt]{geometry}

\Comment{\usepackage{color}
\definecolor{MyDarkBlue}{rgb}{0.15,0.15,0.45}
\usepackage[linktocpage=true]{hyperref}
\hypersetup{
colorlinks=true,
citecolor=MyDarkBlue,\bibliography{myref}
linkcolor=MyDarkBlue,
urlcolor=MyDarkBlue,
pdfauthor={Prieslei Goulart},
pdftitle={DRAFT},
pdfsubject={hep-th}
}

\usepackage[numbers,sort&compress]{natbib}
\usepackage{hypernat}}
\usepackage{graphicx}

\newcommand\ignore[1]{}
\def\one{{\,\hbox{1\kern-.8mm l}}}

\newcommand{\Cset}{{\,\,{{{^{_{\pmb{\mid}}}}\kern-.45em{\mathrm C}}}}}

\newcommand{\be}{\begin{equation}}
\newcommand{\bea}{\begin{eqnarray}}

\newcommand{\ee}{\end{equation}}
\newcommand{\eea}{\end{eqnarray}}

\usepackage{color}
\newcommand{\bse}{\begin{subequations}}
\newcommand{\ese}{\end{subequations}}

\begin{document}

\title{Violation of Weak Cosmic Censorship in Einstein-Maxwell-dilaton theory: singularities connected by traversable wormholes}

\author{Prieslei Goulart}
 \email{prieslei@if.usp.br}
 
\affiliation{
Instituto de F\'{i}sica, Universidade de S{\~a}o Paulo\\ Rua do Mat{\~a}o, 1371, 05508-090 S\~{a}o Paulo, SP, Brazil
}


\begin{abstract}
We give two new analytical solutions to the low-energy string theory action that violate the weak cosmic censorship conjecture. They are classical charged solutions to the Einstein-Maxwell-dilaton theory in four dimensions that come in two types. The first  represents one single naked singularity whose asymptotic region is flat. Its mass respects the positive mass theorem. The absence of horizons allows us to probe quantum aspects of gravity by direct observation of these singularities. The second represents singularities that come in pairs that "seem to create their own spacetime", and the region connecting them has the geometry of a traversable wormhole. In other words, they represent a pair of singularities connected by a traversable wormhole. The dyonic solution has a limit free of singularities, which we call "extremal limit" in analogy with black holes. This has a geometry given by the AdS$_{2}\times$S$^{2}$ spacetime in global coordinates. We compute the physical charges for the naked singularities and show that the weak gravity conjecture can be respected or violated, depending on the parameters of the solutions. We also compute the light deflection angle.
\end{abstract}

\keywords{Cosmic Censorship; Naked Singularities; Exact Solutions}

\maketitle

{\it Introduction.-}The weak cosmic censorship was formulated by Penrose in 1969 \cite{Penrose:1969pc}. It states that singularities can not be naked, {\it i.e.} they must always be covered by a horizon. The cosmic censorship has the status of a hypothesis or conjecture, since it has no formal mathematical proof. It is not an exaggeration to claim that proving or disproving weak cosmic censorship is one of the most important open problems in general relativity. The reason for that is quite simple. It is believed that the singularity can only be described by a quantum theory of gravity. If the phenomena taking place at this singular point of the spacetime result in emission of signals, then, the absence of a horizon would allow a distant observer to receive such signals. In other words, the existence of naked singularities allows direct observation of quantum gravity phenomena. 

In the original formulation \cite{Penrose:1969pc}, Penrose showed that if a spacetime contains a horizon, then it must contain a singularity. However, he did not show that the hypothesis holds true the other way around. Thus, a potential way to violate weak cosmic censorship is to present a physically reasonable solution that contains a singularity which is not covered by a horizon. Nowadays there is strong numerical evidence that weak cosmic censorship is violated in spacetimes with more than four dimensions \cite{Gregory:1993vy,Hubeny:2002xn,Santos:2015iua,Lehner:2010pn,Figueras:2015hkb,Figueras:2017zwa}. Numerical evidence also indicates violation in four dimensions in the Einstein-Maxwell theory with asymptotically anti-de Sitter (AdS) boundary conditions \cite{Horowitz:2016ezu,Crisford:2017zpi}. Although numerics provides strong evidence of violation, a more robust result would be whether a mathematical proof, as in terms of a theorem, or a an analytical solution that provides a naked singularity. In this paper we provide novel analytical solutions to the Einstein-Maxwell-dilaton theory that describe naked singularities. The mass of such singularities is positive. This is a strong indication of stability, although we do not make a stability analysis since this is out of the scope of the paper. The spacetime is static and spherically symmetric, the asymptotic region is flat, and the solutions do not violate causality or null energy condition. As the Einstein-Maxwell-dilaton theory arises as a low-energy effective action in string theory, these are also classical solutions to string theory. 

There are surprising facts about our solutions. The first is that these naked singularities can also exist in pairs, and the region connecting them is described approximately by a traversable wormhole. Since there is no horizon, the singularity at one end interacts with the singularity at the other end. The second is that the dyonic solution has a well defined limit for which the whole spacetime is free of singularities. In this case, the dilaton and the field strengths are constants with respect to the radial coordinate. We will argue that this can not be interpreted as a traversable wormhole, as in the Morris-Thorne sense \cite{Morris:1988cz}. 

{\it Einstein-Maxwell-dilaton theory.-}The fields of the Einstein-Maxwell-dilaton theory are the dilaton $\phi$, the $U(1)$ gauge field $A_{\mu}$, and the metric $g_{\mu\nu}$. The action is written as
\be S=\int d^{4}x\sqrt{-g}\left(R-2\partial_{\mu}\phi\partial^{\mu}\phi-e^{-2\phi}F_{\mu\nu}F^{\mu\nu}\right). \label{ad}\ee
We take units for which $(16\pi G_N)\equiv 1$, where $G_{N}$ is the Newton constant. The Abelian field strength is given by $F_{\mu\nu}=\partial_{\mu}A_{\nu}-\partial_{\nu}A_{\mu}$. The four-dimensional static and spherically symmetric spacetime is given by the metric
\be ds^{2}=-e^{-\lambda(r)}dt^{2}+e^{\lambda(r)}dr^{2}+C^{2}(r)(d\theta^{2}+\sin^{2}\theta d\varphi^{2}),\label{genmet}  \ee
where $\lambda$ and $C$ depend solely on the radial coordinate $r$. The solutions to the gauge field equations (in order to see the equations of motion, see reference \cite{Goulart:2016cuv}) consistent with the Bianchi identities are
\be F_{rt}=\frac{Q}{C^{2}e^{-2\phi}}, \,\,\, F_{\theta\varphi}=P\sin\theta, \ee
where $Q$ and $P$ are constants related to the electric and magnetic charges respectively. The equations of motion of the Einstein-Maxwell-dilaton theory are invariant under S-duality, so one can rotate an electrically charged solution to a magnetically charged one. In the case when the solution to the field equations carry both electric and magnetic charges, we say that we have a dyonic solution, which is invariant under S-duality. 

The black hole solutions of the theory \eqref{ad} were discovered in \cite{Gibbons:1984kp, Gibbons:1987ps}. They were studied from a string theory point of view in \cite{Garfinkle:1990qj}. The multicenter solutions were discovered in \cite{Kallosh:1992ii}, and the massless solution in \cite{Goulart:2016nkv}. In this paper we will present solutions that develop a region in the spacetime where curvature tends to infinity, which gives the location of the singularity. The singularities are naked in the sense that there is no event horizon.

{\it New single charged and dyonic solutions.-}Before presenting the solutions, let us define the following functions
\begin{align}
\alpha_{q}(r) & \equiv \frac{\sqrt{2}k_{q}Q}{l}\arctan\left(\frac{r}{l}\right)+c_{q}, \\
\tilde{\alpha}_{p}(r) & \equiv  \frac{\sqrt{2}k_{p}}{l}\arctan\left(\frac{r}{l}\right)+c_{p}, \\
\alpha_{p}(r) & \equiv \frac{\sqrt{2}k_{p}P}{l}\arctan\left(\frac{r}{l}\right)+c_{p}.
\end{align}
For $P=0$, the Einstein-Maxwell-dilaton theory  \eqref{ad} admits an electrically charged solution given by 
\begin{align}
e^{-\lambda} = & k_{q}\sec\left[ \alpha_{q}(r)\right]  e^{ \tilde{\alpha}_{p}(r)}, \label{ectr}\\
C^{2} = & (r^{2}+l^{2})e^{\lambda}, \label{ecradial}\\
e^{2\phi} = & k_{q}\sec \left[ \alpha_{q}(r)\right]e^{-\tilde{\alpha}_{p}(r)},\label{ecdil}\\
F_{rt} = & \frac{k_{q}^{2}Q}{(r^{2}+l^{2})}\sec^{2} \left[  \alpha_{q}(r)\right].\label{ecef}
\end{align}
Here, $k_{q},\,k_{p}, \, c_{q},\, c_{p}, \,Q$, and $l$ are integration constants, and they must satisfy
\be k_{q}^{2}Q^{2}=l^{2}+k_{p}^2. \label{eccond} \ee
The magnetically charged solution is obtained by applying S-duality transformation. In order to preserve causality, $k_{q}$ must be positive, and the range of the radial coordinate must be chosen such that  
\be -\frac{\pi}{2}<\alpha_{q}(r)<\frac{\pi}{2}. \label{causq} \ee
This implies that the secant function is always positive. We will discuss the existence of singularities in the next section.

The Einstein-Maxwell-dilaton theory \eqref{ad} also admits a dyonic solution given by
\begin{align}
e^{-\lambda} = & k_{q}k_{p}\sec \left[ \alpha_{q}(r)\right]\sec \left[ \alpha_{p}(r)\right], \label{dytr}\\
C^{2} = & (r^{2}+l^{2})e^{\lambda}, \label{dyradial}\\
e^{2\phi} = & \frac{k_{q}\sec \left[ \alpha_{q}(r)\right]}{ k_{p}\sec \left[ \alpha_{p}(r)\right]}, \label{dydil}\\
F_{rt} = & \frac{k_{q}^{2}Q}{(r^{2}+l^{2})}\sec^{2} \left[ \alpha_{q}(r)\right], \,\,\, F_{\theta\varphi}=P\sin\theta.\label{dyefmf}
\end{align}
Here, $k_{q},\, k_{p}, \, c_{q}, \, c_{p},\, Q, \, P$, and $l$ are the integration constants, and they must satisfy the following condition
\be k_{q}^{2}Q^{2}+k_{p}^{2}P^{2}=l^{2}. \label{dycond} \ee
Notice that we do not recover the electrically charged solution directly by setting the magnetic charge $P$ to zero in the dyonic solution: we must also set $k_{p}$ equals to zero in the electrically charged solution. In order to preserve causality, $k_{q}$ and $k_{p}$ must have the same sign (we take only positive signs), and the range of the radial coordinate must be chosen to respect \eqref{causq} and 
\be -\frac{\pi}{2}<\alpha_{p}(r)<\frac{\pi}{2}. \label{causp} \ee
Notice that, without loss of generality, for $c_{q}=c_{p}=0$ and $k_{q}k_{p}=1$, the region around $|r|\approx 0$ is approximately
\be ds^{2}\approx -dt^{2}+dr^{2}+(r^{2}+l^{2})(d\theta^{2}+\sin^{2}\theta d\varphi^{2}).\label{twh} \ee
This is the Bronnikov-Ellis traversable wormhole metric \cite{Fisher:1948yn,Bergmann:1957zza,Bronnikov:1973fh,Ellis:1973yv,Morris:1988cz}, which is a solution to the Einstein-phantom scalar theory. The phantom scalar is the exotic matter because it  violates the null energy condition. In our solutions, the null-energy condition is never violated, even in the wormhole region. Depending on how we choose the integration constants, we might have three different systems: {\it i}) a spacetime that contains one singularity; {\it ii}) a spacetime that contains a pair of singularities; {\it iii}) a spacetime that contains no singularities. 

{\it Curvature and singularities.-}The curvature invariants tell us whether the spacetime has singularities or not. The curvature scalar for the electrically charged and dyonic solutions are respectively
\begin{align}
R^{ec}  = & \frac{k_{q} e^{\tilde{\alpha}_{p}(r)}}{(l^2+r^2)^2} \sec \left[\alpha_{q}(r)\right]\left(k_{q} Q \tan\left[\alpha_{q}(r)\right]-k_{p}\right)^{2},\\
R^{dy} = & \frac{k_{q}k_{p} }{(l^2+r^2)^2} \sec \left[\alpha_{q}(r)\right]\sec \left[\alpha_{p}(r)\right]\left(k_{q} Q \tan\left[\alpha_{q}(r)\right]\right. \nonumber \\
&\left. -k_{p} P \tan \left[\alpha_{p}(r)\right]\right)^{2}. 
\end{align}
They are both positive for the regions defined by \eqref{causq} and \eqref{causp}. They both diverge when the arguments of the tangent functions, {\it i.e.} $\alpha_{q}(r)$ and $\alpha_{p}(r)$, tend to $ -\pi/2$, and to $+\pi/2$. The singularities are located where these divergences occur. The constants $c_{q}$ and $c_{p}$ can be fixed in order to exclude one of the singularities. A spacetime that is completely non-singular is only possible for the dyonic solution.

Due to \eqref{eccond}, the term multiplying the arctangent function in $\alpha_{q}(r)$ is always greater or equal than $\sqrt{2}$ for the electrically charged solution. If it could be less than one, then we would be capable of constructing a spacetime free of singularities representing a traversable wormhole solution, as will be explained later. So, this solution contains at least one and at most two singularities.  Notice that
\be \lim_{r\rightarrow +\infty} \alpha_{q}(r)=\frac{\sqrt{2}k_{q}
Q}{l}\frac{\pi}{2}+c_{q}. \ee
We call the region where $r\rightarrow +\infty$ as the positive asymptotic region, and where $r\rightarrow -\infty$ as the negative asymptotic region. If we want to exclude the singularity at the positive asymptotic region, we must fix $c_{q}$ such that
\be c_{q}<\frac{\pi}{2}\left(1-\frac{\sqrt{2}k_{q}
Q}{l}\right). \label{cond1sing}\ee
We must choose the other integration constants such that \eqref{causq} is still respected. Once we fix $c_{q}$, we also fix the position of the singularity. If we want the spacetime to contain two singularities, it is enough to choose $c_{q}$ in the interval 
\be \frac{\pi}{2}\left(1-\frac{\sqrt{2}k_{q}
Q}{l}\right)\leq c_{q} \leq -\frac{\pi}{2}\left(1-\frac{\sqrt{2}k_{q}
Q}{l}\right). \ee

\begin{figure}[h]
\centering 
\includegraphics[width=9cm, height=7cm]{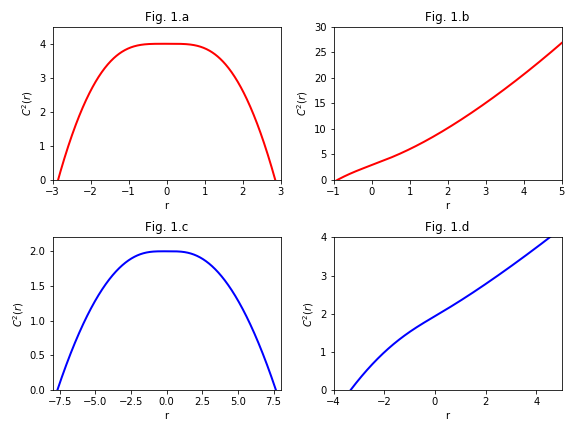}\\
\caption{Solutions with singularities.  Fig. 1.a: Electrically charged pair of singularities for $k_{p}=c_{p}=c_{q}=0$, $k_{q}=0.5$, and $l^{2}=2$. Fig. 1.b: Electrically charged naked singularity for $k_{p}=c_{p}=0$, $c_{q}=0.6 \pi(1-\sqrt{2})$, $k_{q}=0.5$, and $l^{2}=2$. Fig 1.c: Dyonic  pair of singularities for $k_{q}Q_{q}/l=0.8$, $k_{p}P/l=0.6$,  $k_{q}k_{p}=1$, $c_{q}=c_{p}=0$ and $l^{2}=2$. Figure 1.d: Dyonic naked singularity for $k_{q}Q_{q}/l=0.8$, $k_{p}P/l=0.6$,  $k_{q}k_{p}=1$, $c_{q}=0.6 \pi(1-0.8\sqrt{2})$, $c_{p}=0$, and $l^{2}=2$. }\label{fig1}
\end{figure}

For the dyonic solution, the term multiplying the arctangent function can be less than one. Choose for instance $k_{p}P/l<1/\sqrt{2}$. So, for $c_{p}=0$, $-\pi/2<\alpha_{p}(r)<\pi/2$ for all $r$. The singularities are present due to $\alpha_{q}(r)$. Now, because $k_{p}P/l<1/\sqrt{2}$, the condition \eqref{dycond} imposes that $k_{q}Q/l>1/\sqrt{2}$. So, for this choice of parameters we are capable of excluding the singularity at the positive asymptotic region if \eqref{cond1sing} is satisfied for the dyonic solution as well. The singularity in the negative region is always present. For $k_{q}Q/l, k_{p}P/l\neq 1/\sqrt{2}$, and $c_{q}=c_{p}=0$, the spacetime will always contain two singularities, one at each end. As near $|r|\rightarrow 0$ the metric is \eqref{twh}, this configuration represents a pair of singularities connected by a traversable wormhole spacetime. In Fig. \ref{fig1} plot the function $C^{2}(r)$ for these configurations.

There is only one specific choice of parameters for which the dyonic solution has no singularities, which is
\be \frac{k_{q}Q}{l}=\frac{k_{p}P}{l}=\frac{1}{\sqrt{2}}, \,\,\, c_{q}=c_{p}=0. \ee 
This implies that the term multiplying the arctangent function is one. As $-\infty < r < \infty$, the metric and the fields are non-singular in the whole spacetime. In order to discuss this situation, we use the property $\sec[\arctan(x)]=\sqrt{x^{2}+1}$. We also simplify the discussion choosing $k_{q}=e^{\phi_{0}}$ and $k_{p}=e^{-\phi_{0}}$. Notice that the dilaton is a constant field, {\it i.e.}
$ e^{2\phi}=e^{2\phi_{0}}=k_{q}/k_{p}=P/Q$.
This implies that $l=\sqrt{2QP}$, and the field strengths are  independent of $r$, given by $ F_{rt}= 1/2Q, \,\,\, F_{\theta\varphi}=P\sin{\theta}$. Rescaling $t\equiv l\tau$ and $r\equiv l\rho$, the resulting metric is 
\be ds^{2}=l^{2}\left[-(\rho^{2}+1)d\tau^{2}+\frac{d\rho^{2}}{(\rho^{2}+1)}+d\theta^{2}+\sin^{2}\theta d\varphi^{2}\right].\label{global} \ee
This is an AdS$_{2}\times$S$^{2}$ spacetime. The AdS$_{2}$ spacetime is written in terms of global coordinates, $-\infty<\rho<\infty$. The S$^{2}$ sphere has radius $l=\sqrt{2QP}$. 

A classical traversable wormhole corresponds to a spacetime that is completely free of singularities and whose function that multiplies the angular part of the metric, {\it i.e.} $C^{2}(r)$, develops a minimal surface. This minimal surface is called the "throat" of the wormhole. In the Morris-Thorne sense \cite{Morris:1988cz}, $C^{2}(r)$ grows in value away from the throat and the embedding diagram has the form of a paraboloid. The AdS$_{2}\times$S$^{2}$ spacetime \eqref{global} has $C^{2}(r)=l^{2}$, {\it i.e.} a constant. So, this is an infinite cylinder of constant radius, and not really a traversable wormhole according to the above definition.  Notice that the graphical analysis in Fig. \ref{fig1} shows that the solutions with pairs of singularities contain regions for which a traversable wormhole geometry \eqref{twh} is a good approximation, although the function $C^{2}(r)$ develops a maximum there, and not a minimum. 

Wheeler proposed \cite{Wheeler:1955zz} that a pair of particle antiparticle created as a consequence of the Schwinger effect is connected by a wormhole, called nowadays "Wheeler wormhole", which arises due to the same effect. Maldacena and Susskind proposed \cite{Maldacena:2013xja} the so-called ER=EPR conjecture, which is an attempt to relate quantum entanglement with wormholes. In their formulation, the Hawking particles emitted by a black hole could be captured by an outside observer. After some period of time the particles could collapse into a black hole as well. The Hawking radiation is entangled with the black hole that emitted it, so the final system would be a pair of entangled black holes. They also suggested that the association between entanglement and wormholes must be a general feature of all entangled system of particles. In fact, Jensen and Karch \cite{Jensen:2013ora} supported the idea by arguing that an entangled pair of quark and antiquark has the entanglement encoded in a geometry of a non-traversable wormhole in the bulk dual theory. At the same time, Sonner \cite{Sonner:2013mba} showed that such bulk dual corresponds to the Lorentzian continuation of the tunneling instanton describing the Schwinger effect. He also suggested that this is the bulk dual of the creation of a Wheeler wormhole. These proposals are consequences of Schwinger effect and creation of Hawking particles, which are both quantum effects. Following this reasoning , the pair of singularities connected by a traversable wormhole presented here is a classical solution that could very well be interpreted as the bulk dual of a quantum effect in three dimensions. It can not be considered a pair of entangled black holes because the singularities are naked. The fact that the radial coordinate is limited to a finite interval suggests that this pair creates its own spacetime, and only an observer inside the wormhole would be capable of detecting the existence of their fields.

{\it Absence of horizons.-} The position of a horizon for a static metric \eqref{genmet} is found by solving $g_{tt}=e^{-\lambda(r)}=0$ for $r$. This has no solution for \eqref{ectr} and \eqref{dytr}, which shows that our solutions have no horizons.  The positions of the singularity for \eqref{ectr} and \eqref{dytr} with the same value of parameters as in Fig. \ref{fig1} are respectively
\be
r_{S}=-l\tan\left[\left(c_{q}+\frac{\pi}{2}\right)\frac{l}{\sqrt{2}k_{q}Q}\right]\Rightarrow \left\{ \begin{array}{rc} r_{S}^{ec} &  \approx -0.883\\ r_{S}^{dy} & \approx -3.332 \end{array}  \right. . 
\ee
Let us focus our attention to the polar plane $\theta=\pi/2$ and $\varphi=0$. As light travels along the curves defined by $ds^{2}=0$, this implies that the time spent by an object to travel from a position $r_{0}$ to the position of the singularity $r_{S}$ is 
\be c\Delta t=\int_{r_{0}}^{r_{S}} e^{\lambda}dr \Rightarrow \left\{ \begin{array}{rc} \Delta t^{ec} & = 0.82\times 10^7 \text{s},\\ \Delta t^{dy} & = 3.07\times 10^5 \text{s}, \end{array}  \right. \label{timeint} \ee
where we recovered units adding the speed of light $c$ to obtain the numerical values on the right-hand side. The results are finite, which shows the absence of horizons, and are also consistent with the gravitational redshift phenomena which states that clocks tick slower in the presence of gravitational fields.

{\it Conserved charges.-}The pair of singularities does not have an open asymptotic region because the curvature goes to infinity as we approach both ends. So, the asymptotic charges we compute here are related to the naked singularities. Notice that in the positive asymptotic region, the arctangent can be approximated as
\be \arctan\left(\frac{r}{l}\right)\approx \frac{\pi}{2}-\frac{l}{r}. \ee

{Mass:} The $g_{tt}$ component of the electrically charged solution has the following expansion
\begin{align}
e^{-\lambda} \approx & k_{q}\sec(\alpha_{q}^{+})e^{ \frac{\sqrt{2}k_{p}}{l}\frac{\pi}{2}+c_{p}}\nonumber \\
& \times \left(1-\frac{1}{r}\left(\sqrt{2}k_{q}Q\tan(\alpha_{q}^{+})+\sqrt{2}k_{p}\right)\right),
\label{wfec}
\end{align}
where $\alpha_{q}(r\rightarrow +\infty)\equiv \alpha_{q}^{+}$. In the weak field limit, the term multiplying $r^{-1}$ gives the mass of the solution, which implies that the gravitational field can be attractive (positive mass) or repulsive (negative mass). Although our solutions allow for negative mass, this is forbidden by the positive mass theorem \cite{Schon:1979rg,Schon:1981vd, Witten:1981mf}. So, in our paper we will focus on the case when the gravitational field is attractive. Without loss of generality \cite{Note1}, the integration constants can be chosen such that the term multiplying the parenthesis on the right-hand side of \eqref{wfec} is equal to one (fix $c_{q}$, for instance). Then, the mass of the electrically charged solution is
\be
  2M^{\text{ec}}_{+}\equiv \sqrt{2}k_{q}Q\tan(\alpha_{q}^{+})+\sqrt{2}k_{p}. \label{ecposmass}
\ee
The dyonic solution can be approximated as
\begin{align}
e^{-\lambda}\approx & k_{q}k_{p}\sec(\alpha_{q}^{+})\sec(\alpha_{p}^{+})\nonumber \\
& \left(1-\frac{\sqrt{2}k_{q}Q\tan(\alpha_{q}^{+})+\sqrt{2}k_{p}P\tan(\alpha_{p}^{+})}{r}\right),
\label{wfdy}
\end{align}
where $\alpha_{p}(r\rightarrow +\infty)\equiv \alpha_{p}^{+}$.
Again, we can choose the integration constants such that the term multiplying the parenthesis is one, and then the mass of the dyonic naked singularity is
\be
  2M^{\text{dy}}_{+}\equiv \sqrt{2}k_{q}Q\tan(\alpha_{q}^{+})+\sqrt{2}k_{p}P\tan(\alpha_{p}^{+}).\label{dyposmass}
\ee
{Dilaton charge:} The dilaton field has the following expansion in the positive asymptotic region
\be \phi \approx \phi_{+}^{0}-\frac{\Sigma_{+}}{r}+...  \, , \ee
where $\phi_{+}^{0}$ and and $\Sigma_{+}$ are respectively the value of the dilaton and the dilaton charge at $r\rightarrow + \infty$. For the electrically charged naked singularity this gives
\begin{align}
    e^{2\phi_{+}^{0}} &=  k_{q}\sec(\alpha_{q}^{+})e^{-\frac{\sqrt{2}k_{p}}{l}\frac{\pi}{2}-c_{p}}, \\
     2\Sigma_{+} &= -\sqrt{2}k_{q}Q\tan(\alpha_{q}^{+})+\sqrt{2}k_{p},
\end{align}
and for the dyonic naked singularity this gives
\begin{align}
    e^{2\phi_{+}^{0}} &= \frac{ k_{q}\sec(\alpha_{q}^{+})}{k_{p}\sec(\alpha_{p}^{+})}, \\
     2\Sigma_{+} &= -\sqrt{2}k_{q}Q\tan(\alpha_{q}^{+})+\sqrt{2}k_{p}P\tan(\alpha_{p}^{+}).
\end{align}
{Electric charge:} The electric charge $Q_{+}$ is defined by the integral
\begin{equation} 
Q_{+}=\frac{1}{4\pi}\int_{r\rightarrow + \infty}d\theta d\varphi F_{\mu\nu}n^{\mu}m^{\nu}\sqrt{g_{\theta\theta}g_{\phi\phi}},  
\end{equation}
where $m^{\mu}=(1,0,0,0)$ and $n^{\mu}=(0,1,0,0)$. For both  solutions, the electric charge is the parameter $Q$ multiplied by $e^{2\phi_{+}^{0}}$, and due to the fact that the term multiplying the parenthesis in \eqref{wfec} and in \eqref{wfdy} is one, the electric charge is
\begin{equation}
Q_{+}= Qe^{2\phi_{+}^{0}}=k_{q}^{2}Q\sec^{2}{(\alpha_{q}^{+})}.
\end{equation}
{\it Light deflection angle.-}In order to determine the light deflection angle we can use the Post-Parametric-Newtonian (PPN) method \cite{Keeton:2005jd,Keeton:2006sa}. This result is quite general, and gives the light deflection angle with some correction factors for any system with a well defined weak field limit, which is the case of the naked singularities here. The deflection angle is given by
\be \delta_{+} = 4\left(\frac{M_{+}}{b}\right)+\frac{11\pi}{4}\left(\frac{M_{+}}{b}\right)^{2}+32\left(\frac{M_{+}}{b}\right)^{3}. \ee
For both masses \eqref{ecposmass} and \eqref{dyposmass}, $\delta_{+}$ is a positive quantity, which is consistent with an attractive potential.

{\it Comment on weak gravity conjecture.-} The weak gravity conjecture \cite{ArkaniHamed:2006dz} states that any consistent theory of quantum gravity must have a stable particle whose electric charge $q$ is equal or greater than its mass $m$, or just $q/m\geq 1$. It was recently shown \cite{Crisford:2017gsb} that weak gravity conjecture and violation of weak cosmic  censorship in \cite{Horowitz:2016ezu} are incompatible, {\it i.e.} if one is true the other is necessarily false. Equation \eqref{eccond} gives $k_{q}Q =\sqrt{l^{2}+k_{p}^{2}}$, so, the ratio between the electric charge and the mass is
\be \frac{Q_{+}}{M^{ec}_{+}}= \sqrt{2}k_{q}\left[\frac{\sec^{2}(\alpha_{q}^{+})}{k_{p}+\sqrt{l^{2}+k_{p}^{2}}\tan\left(\alpha_{q}^{+}\right)}\right]. \ee
We can make $k_{q}$ as small as we wish while keeping the product $k_{q}Q$ fixed. This also keeps $\alpha_{q}^{+}$ fixed. In this situation, $Q_{+}/M^{ec}_{+}\rightarrow 0$. We can also make $k_{q}$ as big as we wish, keep the product $k_{q}Q$ fixed, and obtain $Q_{+}/M^{ec}_{+}\rightarrow +\infty$. So the weak gravity conjecture can be respected or violated by the electrically charged naked singularity. Perhaps, stability analysis imposes extra restrictions on the allowed values for the parameters such that there is no violation of weak gravity conjecture.

{\it Discussion.-}There are several consequences related to the analytical solutions we presented and we remark some here: {\it i.} The weak cosmic censorship no longer holds true in four-dimensional asymptotically flat spacetime. This proves to be very important since we can observe directly quantum gravity phenomena. If the naked singularity is observed in nature and is described by our solution, then this will support string theory as a consistent theory of quantum gravity. {\it ii.} It is possible to obtain theoretically charged pairs connected by wormholes without the need of quantum mechanical effects. {\it iii.} Weak gravity conjecture does not play an important role in our solutions. {\it iv.} The AdS$_{2}\times$S$^{2}$ spacetime arises as a sort of extremal limit of the dyonic solution.

There are many open questions to be answered. The most immediate one is whether our solutions are stable under small perturbations of the metric, which is a topic of future work. There is a possibility that naked singularities are formed in stellar collapse. The Event Horizon Telescope will observe regions of the space where stars seem to orbit some object that is not observed in the optical spectrum. If these are black holes, then it will be possible to construct a picture of the their event horizon. If the event horizon is absent then this is a strong indication that the stars may be orbiting a naked singularity. If event horizons are present, then our solutions are still not ruled out, since they might represent the final stage of evaporating black holes.\\
{\it Acknowledgements.-}PG is grateful to Horatiu Nastase and Victor Rivelles for comments on the manuscript. This work is supported by FAPESP grant 2017/19046-1.



\providecommand{\href}[2]{#2}\begingroup\raggedright
\endgroup
\end{document}